\documentclass[%
 reprint,
superscriptaddress,
 amsmath,amssymb,
 aps,
prb,
]{revtex4-2}

\usepackage{graphicx}
\usepackage{dcolumn}
\usepackage{bm}

\usepackage{color}
\usepackage{textcomp}
\usepackage{mathcomp}
\usepackage{multirow}

\begin{document}


\title{\textbf {
Investigation of the Paramagnetic State of the Kagome Kondo Lattice Compound YbV$_6$Sn$_6$: a $^{51}$V Nuclear Magnetic Resonance Study
} 
}%

\author{S. Park}
\affiliation{Materials Physics and Applications–Quantum, Los Alamos National Laboratory, Los Alamos, New Mexico 87545, USA}

\author{H. Sakai}
\email{sakai.hironori@jaea.go.jp}
\affiliation{Advanced Science Research Center, Japan Atomic Energy Agency, Tokai, Ibaraki 319-1195, Japan}

\author{S. Hosoi}
\affiliation{Materials Physics and Applications–Quantum, Los Alamos National Laboratory, Los Alamos, New Mexico 87545, USA}

\author{S. M. Thomas}
\affiliation{Materials Physics and Applications–Quantum, Los Alamos National Laboratory, Los Alamos, New Mexico 87545, USA}

\author{S. Kambe}
\affiliation{Advanced Science Research Center, Japan Atomic Energy Agency, Tokai, Ibaraki 319-1195, Japan}

\author{Y. Tokunaga}
\affiliation{Advanced Science Research Center, Japan Atomic Energy Agency, Tokai, Ibaraki 319-1195, Japan}

\author{A. P. Dioguardi}
\affiliation{Materials Physics and Applications–Quantum, Los Alamos National Laboratory, Los Alamos, New Mexico 87545, USA}

\author{J. D. Thompson}
\affiliation{Materials Physics and Applications–Quantum, Los Alamos National Laboratory, Los Alamos, New Mexico 87545, USA}

\author{F. Ronning}
\affiliation{Materials Physics and Applications–Quantum, Los Alamos National Laboratory, Los Alamos, New Mexico 87545, USA}

\author{M. Kimata}
\affiliation{Advanced Science Research Center, Japan Atomic Energy Agency, Tokai, Ibaraki 319-1195, Japan}
\affiliation{Institude for Materials Research, Tohoku University, Sendai, Miyagi 980-8577, Japan}

\author{T. Furukawa}
\affiliation{Institude for Materials Research, Tohoku University, Sendai, Miyagi 980-8577, Japan}

\author{T. Sasaki}
\affiliation{Institude for Materials Research, Tohoku University, Sendai, Miyagi 980-8577, Japan}

\author{E. D. Bauer}
\affiliation{Materials Physics and Applications–Quantum, Los Alamos National Laboratory, Los Alamos, New Mexico 87545, USA}

\author{M. Hirata}
\email{mhirata@lanl.gov}
\affiliation{Materials Physics and Applications–Quantum, Los Alamos National Laboratory, Los Alamos, New Mexico 87545, USA}

\date{\today}

\begin{abstract}

YbV$_6$Sn$_6$ is a recently discovered kagome-lattice metal that orders at $T_{\rm N}\approx0.4$~K.
Its layered structure combines a triangular Kondo lattice of Yb$^{3+}$ ions with vanadium-based kagome planes, which may host an interplay between strong correlations and band topology.
We report a $^{51}$V nuclear magnetic resonance (NMR) study of the paramagnetic state of YbV$_6$Sn$_6$.
Detailed field-angular dependence of single-crystal NMR spectra determined the principal-axis directions of the electric field gradient tensor at the $^{51}$V sites, as well as their nuclear quadrupole frequency, $\nu_{\rm Q}$, and asymmetry parameter, $\eta$.
The Knight shift, $K$, was measured for different field orientations, and the analysis of $K$ against magnetic susceptibility to extract anisotropic hyperfine couplings.
Accurate spectral assignments further enabled measurements of the nuclear spin-lattice relaxation rate, $1/T_1$, for both in-plane and out-of-plane field directions.
The temperature dependence of $1/T_1$ shows that out-of-plane spin fluctuations are suppressed below $\sim$20~K, whereas in-plane fluctuations are markedly enhanced, which might be understood by thermal depopulation of the low-lying crystalline electric field excited state.
The notable anisotropy in $1/T_1$ indicates that
the paramagnetic state of YbV$_6$Sn$_6$ is strongly affected by in-plane spin dynamics.
\end{abstract}

\maketitle


\section{\label{sec:introduction}Introduction}

The kagome lattice possessing corner-sharing triangular geometry has long served as a fertile platform for exploring
emergent phenomena linked to geometrical frustration, particularly in localized spin systems.
Magnetic
order is often suppressed in such a lattice due to frustration, making the kagome system a prime candidate for hosting quantum-spin-liquid states \cite{Balents2010Spin-liquids-in,Broholm2020Quantum-spin-li} as well as other unconventional magnetic ground states \cite{Messio2012Kagome-Antiferr,Mondal2021Regular-magneti}.
More recently, itinerant electron systems on kagome lattices have attracted growing interest, where the interplay of band topology, electron correlations, and lattice geometry gives rise to a variety of exotic quantum phenomena \cite{Wang2024Topological-Qua}.
In particular, metallic vanadium-based kagome compounds exhibit unconventional orders such as charge density wave (CDW) and superconductivity, often driven by singular features of kagome band structures near the Fermi level.
Representative examples include $A$V$_3$Sb$_5$ ($A$ = K, Rb, Cs) \cite{Ortiz2019New-kagome-prot, Ortiz2021Superconductivi, jiang2021unconventional-} and $AE$V$_3$Sb$_4$ ($AE$ = Ca, Yb, Eu) \cite{Ovchinnikov2020Bismuth-as-a-Re, Ortiz2023YbV3Sb4-and-EuV} families, which exhibit unconventional CDW order \cite{Wilson2024AV3Sb5-kagome-s,Jiang2023Kagome-supercon,jiang2021unconventional-,Wang2021Electronic-natu} and superconductivity \cite{Wilson2024AV3Sb5-kagome-s,Wu2021Nature-of-Uncon,Jiang2023Kagome-supercon}, stimulating intensive studies of their nontrivial ground states. 

In a broader context of strongly correlated electron systems with 4$f$ or 5$f$ electrons, the interplay between Kondo coupling and magnetic frustration has emerged as a key aspect for realizing novel quantum phases \cite{Si2006Global-magnetic, Vojta2008From-itinerant-, Coleman2010Frustration-and,Batista2016Frustration-and}.
In Kondo lattice systems with a periodic array of $f$-electron moments, the localized $f$-electrons interact with conduction electrons, often leading to heavy-fermion behavior and unconventional ground states.
Geometrical frustration in such lattices has been proposed to further enhance quantum fluctuations and potentially give rise to exotic phases, such as quantum spin liquids \cite{Burdin2002Heavy-fermion-a} or partially ordered states \cite{Motome2010Partial-Kondo-S}.

YbV$_6$Sn$_6$ is a recently found material belonging to the $R$Mn$_6$Sn$_6$-type 166 structure ($R$ = rare earth) that offers a rare example of a triangular Kondo lattice formed by ytterbium ions~\cite{Guo2023Triangular-Kond}.
In the Yb trigonal lattice, Yb$^{3+}$ ions carry a moment with effective spin $J_{\text{eff}} = 1/2$ in the crystalline electric field (CEF) ground-state doublet \cite{Li2015Rare-Earth-Tria,Schmidt2021Yb-delafossites,Bag2021Realization-of-}, with a low-lying CEF excited state showing up $\Delta/k_{\rm B} \approx$~30~K above the ground state \cite{Guo2023Triangular-Kond}.
A magnetic transition from a paramagnetic state to a putative antiferromagnetic (AFM) state has been reported at $T_{\rm N} \approx  0.4$~K \cite{Guo2023Triangular-Kond}, and characteristic heavy-fermion behavior is found as reflected in a sizable residual Sommerfeld specific heat coefficient of 411~mJ mol$^{-1}$K$^{-2}$~\cite{Guo2023Triangular-Kond}. 
Notably, the triangular lattice of Yb makes an array of rather localized 4$f$-electron moments that is structurally sandwiched by metallic vanadium kagome layers.
This layered structure makes YbV$_6$Sn$_6$ a compelling platform that combines Kondo physics with the metallic kagome network, offering an opportunity to explore the coupling between frustrated local moments and itinerant electronic states.

Nuclear magnetic resonance (NMR) is a powerful local probe that offers micro-scale access to electronic and magnetic correlations from a site-selective spectroscopy.
In particular, $^{51}$V NMR is well suited for such studies, as the $^{51}$V nucleus has a nucleat quadrupole moment that couples to the local electric field gradient (EFG).
This coupling provides valuable information about the local site symmetry and charge distributions, which complements the magnetic insights
acquired from Knight shift measurements.

In this study, we present a detailed $^{51}$V NMR investigation of YbV$_6$Sn$_6$ in its paramagnetic state.
Our results provide microscopic insights into the local electronic and magnetic environment of this compound, serving as a reference for future NMR studies of vanadium-based kagome materials.

\section{\label{sec:experimental}Experimental Details}

Single crystals of YbV$_6$Sn$_6$ were grown from tin flux, similar to a previous report \cite{Guo2023Triangular-Kond}.
Yb, V, and Sn were placed in an alumina crucible in the ratio 1:6:30, and sealed in a quartz tube under vacuum.
The tube was heated to 1175$^\circ$C, held there for 18 hours, then cooled slowly to 750$^\circ$C, where the Sn flux was removed by centrifugation.  The crystals were etched in HCl to remove Sn on the surface.
The crystals often formed as hexagonal plates of order 1-3 mm on a side and ~0.1-0.2 mm thick.
Lattice parameters and magnetic susceptibility (measured from 1.8 to 300~K under $\mu_0 H_0 = 3$~T using a Quantum Design, MPMS) were consistent with Ref. \onlinecite{Guo2023Triangular-Kond}.
In addition, AC magnetic
susceptibility measurements were performed using a dilution
refrigerator (Oxford, Proteox) to confirm a phase transition at $T_{\mathrm{N}} \approx 0.4$~K \cite{Guo2023Triangular-Kond}.

For NMR experiments, single-crystalline samples of YbV$_6$Sn$_6$ were loaded in a standard $^{4}$He superconducting magnet and variable temperature insert cryostat.
Each specimen was mounted on a glass slip and wrapped with a silver wire to form a radio frequency (rf) coil. 
Dual-axis goniometers were used to make an in-situ rotation of the sample coil in an external magnetic field.
The NMR measurements at $^{51}$V nuclei (nuclear spin $I=7/2$) were carried out using a spin-echo technique with a standard phase-coherent spectrometer.
NMR spectrum was obtained through a Fast-Fourier-Transform (FFT) of the spin echo signal under varying carrier frequencies.
The nuclear quadrupole frequency is defined as
$\nu_{\rm Q}\equiv \left| 3e^{2}qQ / \{2I(2I-1)h\} \right|$, where $eQ$ is the nuclear quadrupole moment, $I$ is the nuclear spin quantum number, 
and $eq\equiv V_{ZZ}$ is the largest principal component of the EFG tensor.
Here, $V_{ii}$ is defined as the principal component in the $i$-axis direction for each $^{51}$V nuclei (with $|V_{XX}|\le |V_{YY}|\le |V_{ZZ}|$ and $V_{XX}+V_{YY}+V_{ZZ}=0$).
The asymmetry parameter is given by $\eta\equiv {|V_{YY}-V_{XX}|}/{|V_{ZZ}|}$.
We used the nuclear quadrupolar moment for $^{51}$V of $^{51}Q=-0.052 \times 10^{-24}$ cm$^2$ \cite{Pyykko2018Year-2017-nucle}
and the nuclear gyromagnetic ratio $^{51}\gamma_{\rm n}/2\pi= 11.193$ MHz/T.

The nuclear spin-lattice relaxation time, $T_1$, was measured using an inversion-recovery method.
Values of $T_{1}$ were obtained from fits to an appropriate relaxation function.
The magnetization recovery ($\{ M(\infty)-M(t)\} /M(\infty)$) for the third quadrupole satellite peak of the $^{51}$V ($I=7/2$) NMR spectrum gave satisfactory fits to the single-$T_{1}$ functions:
$\frac{1}{84}\exp(-t/T_1)+\frac{3}{28}\exp(-3t/T_1)+\frac{3}{11}\exp(-6t/T_1)+\frac{25}{77}\exp(-10t/T_1)+\frac{75}{364}\exp(-15t/T_1)+\frac{3}{44}\exp(-21t/T_1)+\frac{4}{429}\exp(-28t/T_1)$.

\section{\label{sec:results}Results and discussions}

\subsection{Frequency-swept $^{51}$V NMR spectra and site assignments}

\begin{figure}[tb]
\includegraphics[width=8.5cm]{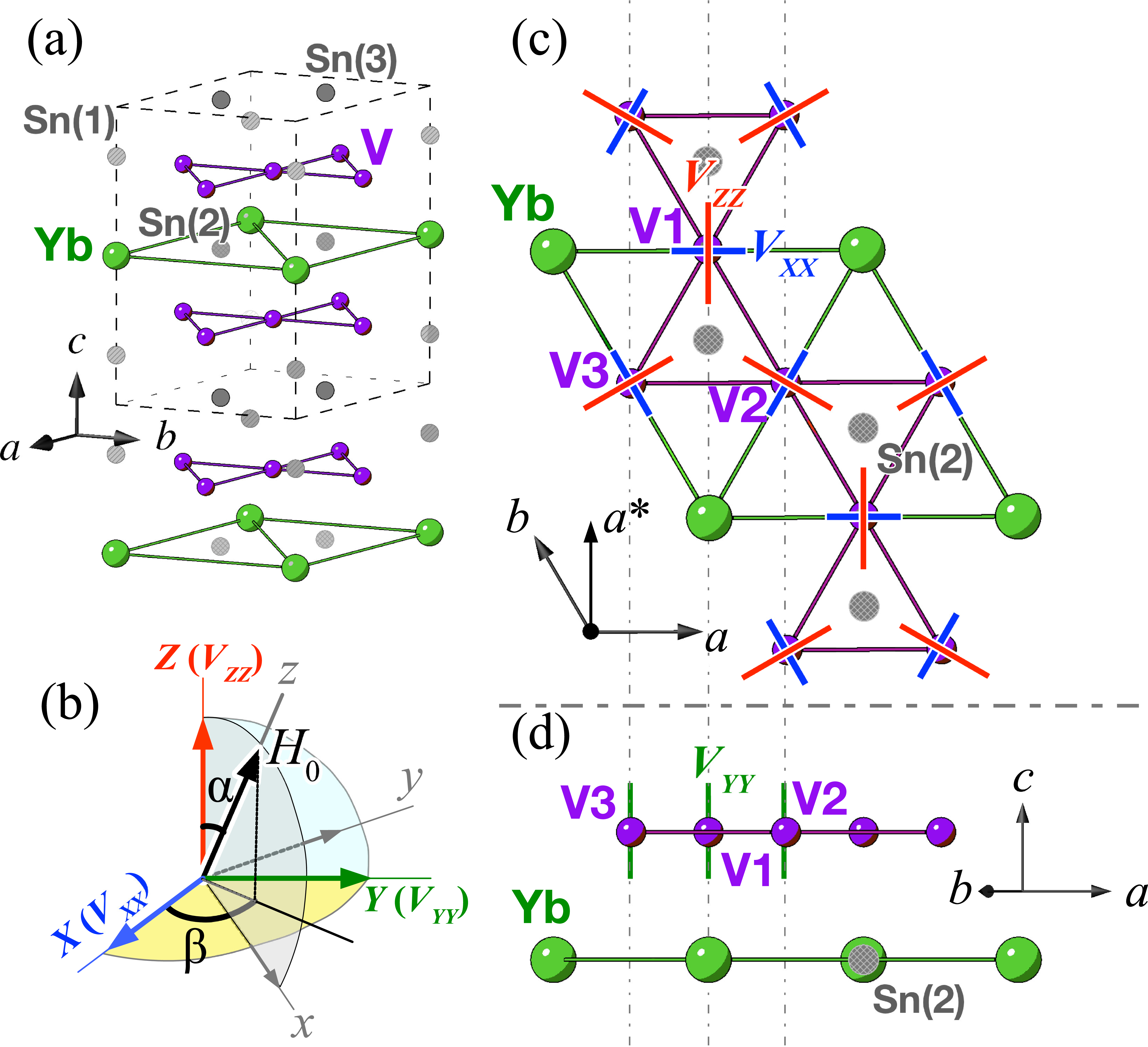}
\caption{\label{fig:CrystalStructure}
(a) Crystal structure of YbV$_6$Sn$_6$. The unit cell is shown by the dashed lines, with Yb, V, and Sn atoms given by green, purple, and gray balls, respectively. (b) Local frame of an external magnetic field, $H_0$, given in a spherical coordinate relative to the principal axes of the electric field gradient (EFG) tensor at V sites. (The $xyz$-coordinate represents the laboratory frame, with $z$ pointing to the direction of $H_0$.) The polar angle $\alpha$ is formed between $H_0$ and the $V_{ZZ}$-axis, and the azimuthal angle $\beta$ is measured from the $V_{XX}$-axis. (c) and (d): Top view (c) and side view (d) of the crystal structure seen parallel and perpendicular to the $c$ axis, respectively. Principal axes of the EFG tensor at each V site are illustrated by blue ($V_{XX}$), green ($V_{YY}$), and red ($V_{ZZ}$) bars. Note that a crystallographically unique V site triples into V1, V2, and V3 sites when $H_0$ is applied (see the main text). 
}
\end{figure}

\begin{figure}[htb]
\includegraphics[width=8.5cm]{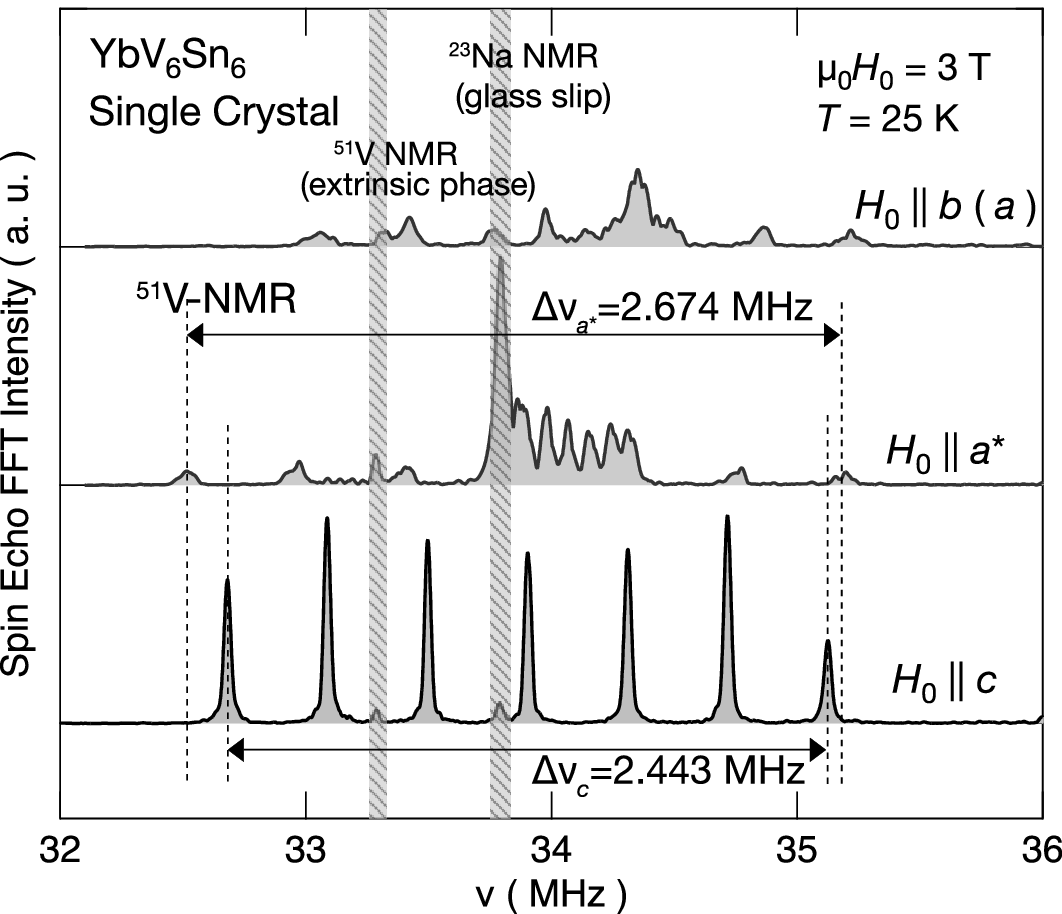}
\caption{\label{fig:NMRspectra} Frequency-swept $^{51}$V NMR spectra of single crystal YbV$_6$Sn$_6$ measured at $T = 25$ K and $\mu_0 H_0=$~3~T applied along the crystalline $b$~axis (top), $a^*$~axis (middle), and $c$~axis (bottom).
The spacing between the outermost satellite lines is given by double ended arrows for $H_0 \parallel a^*$ ($\Delta \nu_{a^*}$) and $H_0 \parallel c$ ($\Delta \nu_c$) (see the text). The shaded vertical bars indicate extrinsic signals from $^{51}$V in minor impurity phases and $^{23}$Na in the glass slip used for the sample mount.}
\end{figure}

\begin{figure*}[hbt]
\includegraphics[height=9cm]{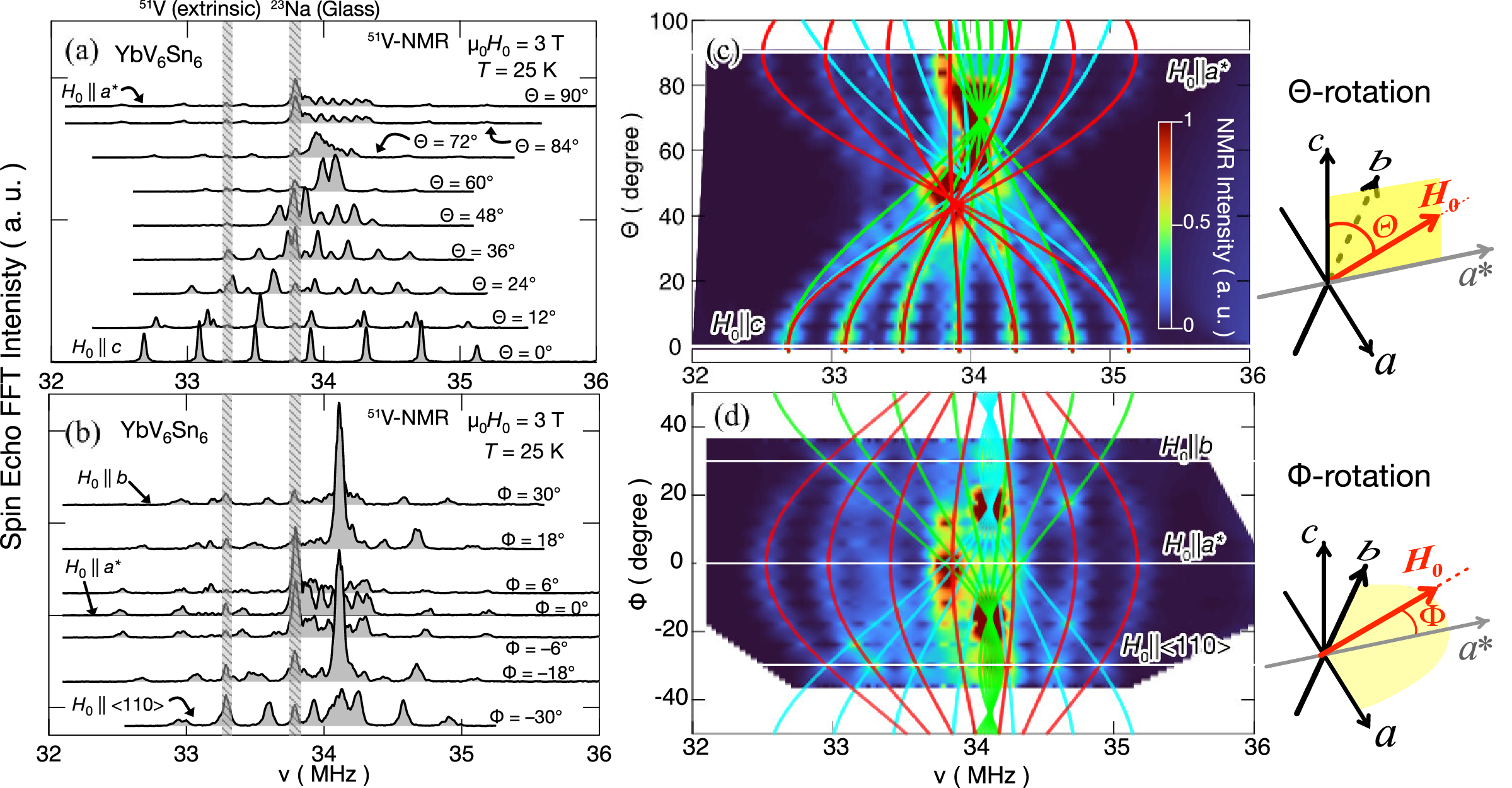}
\caption{\label{fig:AngularDepSpectra} Field-orientation dependence of frequency-swept $^{51}$V NMR spectra of single-crystal YbV$_6$Sn$_6$ measured at $T = 25$~K and $\mu_0H_0 = 3$~T. The magnetic field $H_0$ is rotated within the (a) $a^\ast c$-plane ($\Theta$-rotation) and (b) $ab$-plane ($\Phi$-rotation). In panels (a) and (b), representative spectra at selected angles are shown. The field orientation is defined by the angle $\Theta$ from the $c$ axis in (a), and by the angle $\Phi$ from the $a^\ast$ axis in (b). Panels (c) and (d) present color contour plots of the full angular dependence of signal intensity, overlaid with simulated resonance frequencies (solid curves) corresponding to (a) and (b), respectively. The diagrams to the right of (c) and (d) schematically illustrate the angle definitions relative to the crystal axes. Red, green, and blue curves indicate the seven quadrupole-split NMR lines from the three inequivalent V sites [V1, V2, and V3 in Fig.~\ref{fig:CrystalStructure}(c) and (d)]. Gray-shaded bars in (a) and (b) indicate extrinsic signals that show no angular dependence. }
\end{figure*}

The crystal structure of YbV$_6$Sn$_6$ with the space group $P6/mmm$ (\#191) is shown in Fig.~\ref{fig:CrystalStructure}(a).
Crystallographically, the structure contains one V site, one Yb site, and three crystallographically distinct Sn sites.
The V atoms form a kagome lattice within the $ab$-plane, while Yb atoms form triangular layers sandwiched between V layers, resulting in a stacking sequence along the $c$ direction of $\cdots$–V–V–Yb–V–V–Yb–$\cdots$.

The V site corresponds to the Wyckoff position $6i$ and a local site symmetry $2mm$ with an orthorhombic local environment.
Under an external magnetic field $H_0$, the three V sites become magnetically nonequivalent (dubbed V1, V2, and V3 hereafter), depending on the field orientation relative to the principal axes of the EFG tensor that are unique to each of the three $^{51}$V sites.
(This magnetic nonequivalence is reflected in the observation of more-than-seven NMR lines as discussed below.)
In Fig.~\ref{fig:CrystalStructure}(c) and (d) we illustrate the principal axes of the EFG tensor at V1, V2, and V3 sites, and 
Fig.~\ref{fig:CrystalStructure}(b) relates $H_0$ in the laboratory ($xyz$) frame to the principal-axis ($XYZ$) frame of the EFG tensor for one of the three V sites.

Figure~\ref{fig:NMRspectra} shows representative $^{51}$V NMR spectra measured at $T=$~25~K for $\mu_0H_0= 3$ T applied along the crystalline $b$, $a^\ast$, and $c$ axis, by accurately aligning the sample using a dual-axis goniometer.
Since $^{51}$V has nuclear spin $I = 7/2$, it experiences quadrupole interactions, and a unique V site gives rise to seven quadrupole-split NMR lines: one central line, three satellites on the lower-frequency side [L1, L2, and L3], and three on the higher-frequency side [H1, H2, and H3], with the satellite lines numbered outward from the central line.
For $H_0 \parallel c$, only seven lines appear with an even spacing, indicating that all V sites are magnetically equivalent.
This also implies that the crystalline $c$ axis is aligned with at least one of the three principal axes of the EFG tensor for all V sites.
By contrast, for $H_0 \parallel a^\ast$ and $b$, more than seven peaks are observed, as the unique V environment for $H_0 \parallel c$ turns into multiple nonequivalent ones when the magnetic field is tilted off the $c$ axis.

The nuclear spin Hamiltonian consists of nuclear Zeeman ($\mathcal{H}_{\rm Z}$) and quadrupolar ($\mathcal{H}_{\rm Q}$) terms: 
\begin{equation}
	\mathcal{H}_{\rm Z}=\gamma_{\rm n}\hbar\{1+K(\alpha, \beta)\} \bm{I}\cdot\bm{H_0},
	\label{eq:Hz}
\end{equation}

\noindent and 
\begin{equation}
\mathcal{H}_{\rm Q}=\frac{h\nu_{\rm Q}}{6} \Big\{3I_{z}^{2}-I(I+1)+\frac{\eta}{2}(I_{+}^{2}+I_{-}^{2}) \Bigr\},
\label{eq:Hq}
\end{equation}
\noindent with the total Hamiltonian $\mathcal{H} = \mathcal{H}_{\rm Z} + \mathcal{H}_{\rm Q}$.

When the quadrupole term is much smaller than the Zeeman term ($\mathcal{H}_{\rm Q} \ll \mathcal{H}_{\rm Z}$), a perturbative approach is useful for extracting the EFG parameters of $\nu_{\rm Q}$ and $\eta$.
In this limit, within the first-order perturbation, the outermost spacing between the satellites L3 and H3 is related to the EFG parameters as
\begin{multline}
\Delta\nu(\alpha, \beta)
= 3\nu_{\rm Q} \left| 3\cos^2\alpha - 1 + \eta\sin^2\alpha\cos 2\beta \right|,
\label{eq:SatelliteInterval}
\end{multline}
which is valid regardless of the Knight shifts and second-order perturbation.
The maximum satellite spacing is realized for $\alpha = 0$ (i.e., $H_0 \parallel V_{ZZ}$).

As depicted in Fig.~\ref{fig:NMRspectra}, such maximized $\Delta\nu$ seems to appear for $H_0 \parallel a^\ast$ and not $H_0 \parallel c$, suggesting that $V_{ZZ}$ may lie within the $ab$-plane (rather than along the $c$ axis).
To verify this interpretation, it is necessary to assign the full NMR spectra to the V1, V2, and V3 sites and identify the correct pair of L3 and H3 satellite peaks for determining the quadrupolar splitting.

\subsection{EFG principal axes revealed by angular-dependent NMR spectroscopy}\label{sec:EFG}

To determine the directions of the principal axes of the EFG tensor for the three V sites, we examined the field-orientation dependence of the $^{51}$V NMR spectra by rotating $H_0$ within the $a^\ast c$-plane ($\Theta$-rotation; angle $\Theta$ measured from $c$ axis) and the $ab$-plane ($\Phi$-rotation; angle $\Phi$ measured from $a^*$ axis).
Figure~\ref{fig:AngularDepSpectra}(a) and \ref{fig:AngularDepSpectra}(b) present angular dependence of frequency-swept spectra for $\Theta$ and $\Phi$ rotations, respectively, and Fig.~\ref{fig:AngularDepSpectra}(c) and \ref{fig:AngularDepSpectra}(d) give the two-dimensional color contour plots of the signal intensity for the corresponding rotations, respectively.
First, the angular dependence of satellite-peak positions was carefully traced in the $\Theta$-rotation, which allowed us to assign the full spectra to signals from three different V sites [Fig.~\ref{fig:AngularDepSpectra}(a) and \ref{fig:AngularDepSpectra}(c)]. The spacing $\Delta\nu$ between L3 and H3 for one V site was found to become largest for $H_0 \parallel a^\ast$, suggesting that $V_{ZZ}$ for that site (defined as V1) lies along the $a^\ast$ direction.
This assignment is further supported by the $\Phi$-rotation, where the satellite spacing $\Delta\nu$ for V1 site is confirmed to be maximized for $H_0 \parallel a^*$ [Fig.~\ref{fig:AngularDepSpectra}(b) and \ref{fig:AngularDepSpectra}(d)].
This leads us to conclude that the $c$ axis corresponds to the $V_{YY}$ direction for all three V sites and the remaining $V_{XX}$ axis for the V1 site must lie along the $a$ direction.

Consequently, for the V1 site under $H_0 \parallel a^*$, the satellite spacing is given by $\Delta\nu(0, \frac{\pi}{2}) = 6\nu_{\rm Q}$ according to Eq.~(\ref{eq:SatelliteInterval}).
Similarly, for $H_0 \parallel c$, the spacing becomes $\Delta\nu(\frac{\pi}{2}, \frac{\pi}{2}) = 3\nu_{\rm Q}(1 + \eta)$.
From these relations, we obtain $\nu_{\rm Q} = 0.446 \pm 0.001$~MHz and $\eta = 0.826 \pm 0.001$.
The principal axes for V2 and V3 sites can be determined from symmetry arguments as illustrated in Fig.~\ref{fig:CrystalStructure}(c) and \ref{fig:CrystalStructure}(d), reflecting the crystal’s sixfold rotational symmetry around the $c$ axis.

To validate the experimentally determined EFG parameters, we performed density functional theory (DFT) calculations using the Perdew–Burke–Ernzerhof exchange-correlation functional \cite{PerdewJP:PRL77:1996}  implemented in WIEN2k \cite{LuitzPBKSGMDKJ:TUWA:2001}.
Spin–orbit coupling was included in a second-variational scheme without relativistic local orbitals.
The calculated values, $\nu_{\rm Q} = 0.458$ MHz and $\eta = 0.92$, are in good agreement with experiment.

The three Knight shift components $K_i$ ($i = a, a^\ast$, and $c$) are defined in the crystallographic frame and transformed into the local EFG principal-axis system for each V site.
For example, in the case of $\Theta$-rotation, the angular dependence of the Knight shift is expressed as:
$K(\alpha, \beta) = K(\frac{\pi}{2}-\Theta, \frac{\pi}{2}) = K_c \sin^2\Theta + K_{a^\ast} \cos^2\Theta$ for V1;
$K(\alpha, \beta) = K(\frac{\pi}{2}-\frac{2}{3}\Theta, \frac{\pi}{2}-\Theta)$ for V2; and
$K(\alpha, \beta) = K(\frac{\pi}{2}-\frac{2}{3}\Theta, \frac{\pi}{2}-\Theta)$ for V3.
A similar transformation is applied for the $\Phi$-rotation. 
Using these $K(\alpha, \beta)$, we simulated the spectra and fit to the angular dependence of peak positions through exact diagonalization of the total Hamiltonian $\mathcal{H} = \mathcal{H}_{\rm Z} + \mathcal{H}_{\rm Q}$ [as shown by solid curves in Fig.~\ref{fig:AngularDepSpectra}(c) and \ref{fig:AngularDepSpectra}(d)].
Employing $K_a = 1.4 \pm 0.1$\%, $K_{a^\ast} = 0.8 \pm 0.1$\%, and $K_c = 0.5 \pm 0.05$\%, the simulation successfully reproduced all the patterns of experimental angular dependence in Fig.~\ref{fig:AngularDepSpectra}.

\subsection{Temperature dependence of Knight shift and magnetic susceptibility}\label{sec:KnightShift}

\begin{figure}[hbt]
\includegraphics[width=8.5cm]{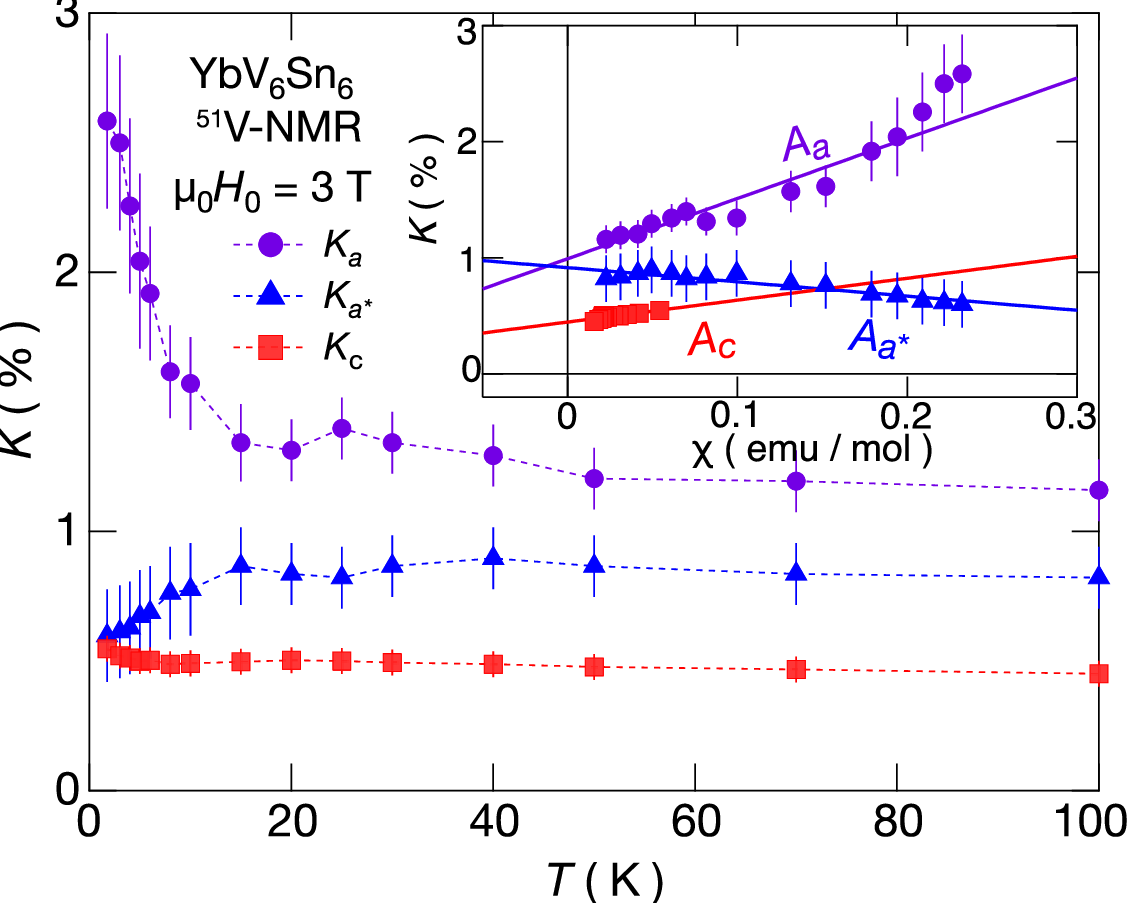}
\caption{\label{fig:KT_Kchi} Temperature dependence of the $^{51}$V Knight shifts in three different field orientations: $K_a$, $K_{a^\ast}$, and $K_c$, measured at $\mu_0 H_0 = 3$~T.
The inset shows a $K$–$\chi$ plot for the corresponding field directions, with solid lines representing least-squares fits.
The magnetic susceptibility $\chi$ is measured at 3~T on another sample from the same batch and is defined as the component perpendicular ($\chi_{\perp}$) or parallel ($\chi_{\parallel}$) to the $c$ axis.}
\end{figure}

The three shift values at different temperatures are determined by fitting the full spectra of the V1, V2, and V3 sites using exact diagonalization of the nuclear spin Hamiltonian $\mathcal{H}$, assuming that both $\nu_{\rm Q}$ and $\eta$ are temperature-independent.
The $\nu_{\rm Q}$ and $\eta$ were checked not to show any notable change below 100 K.
To reduce the number of fitting parameters, we fixed them as constants.
Figure~\ref{fig:KT_Kchi} shows the temperature dependence of the $^{51}$V Knight shift $K$ for three different field orientations of $H_0 \parallel a$~($K_a$), $H_0 \parallel a^*$~($K_{a^\ast}$), and $H_0 \parallel c$~($K_c$).
A strong easy-plane anisotropy is observed in $K$, especially below $\sim$~20~K, consistent with the earlier magnetic susceptibility $\chi$ measurements \cite{Guo2023Triangular-Kond}.  

Among these, $K_c$ has the highest accuracy (smallest error bars) because it is determined from the well-resolved seven-peak spectra for $H_0 \parallel c$ (bottom plot in Fig.~\ref{fig:NMRspectra}).
In contrast, $K_a$ and $K_{a^\ast}$ have larger uncertainties, especially at low $T$, due to a spectral broadening and signal overlap from magnetically nonequivalent V sites [V1, V2, and V3 in Fig.~\ref{fig:CrystalStructure}(c) and \ref{fig:CrystalStructure}(d)].
These uncertainties are reflected in the error bars shown in Fig.~\ref{fig:KT_Kchi}.

Standard $K$–$\chi$ analyses are performed for each field direction to estimate the hyperfine coupling constant, using temperature as an implicit parameter (inset of Fig.~\ref{fig:KT_Kchi}).
Here, $K_c$ and $K_a$ (and $K_{a^*}$) are plotted against out-of-plane susceptibility ($\chi_\parallel$) and in-plane susceptibility ($\chi_{\perp}$), respectively, both measured under the same magnetic field of 3 T using another sample from the same batch.
Given that error bars are comparable to scatter in the data, particularly at low $T$, it is reasonable to assume a linear relation between $K$ and $\chi$ over the entire $T$ range for all three directions, with a finite intercept, $K_0$, observed at $\chi = 0$.
Although $\chi(T)$ follows a Curie–Weiss law and the $T$-independent term $\chi_0$ is nearly zero, the finite $K_0$ seen here suggests that there may be a finite metallic Pauli contribution and/or van-Vleck paramagnetism of Yb$^{3+}$ ions transferred to V nuclei via hyperfine coupling.
Such finite $K_0$ values are commonly observed in strongly correlated $f$-electron systems, such as the U$T$Ga$_5$ ($T$ = Pt, Ni) compounds \cite{Haru_U115_JPSJ}.
From the slopes of three $K$–$\chi$ lines, we determine highly anisotropic hyperfine coupling constants: $A_a = 29 \pm 4$~mT/$\mu_{\rm B}$, $A_{a^\ast} = -7 \pm 3$~mT/$\mu_{\rm B}$, and $A_c = 10.5 \pm 1.5$~mT/$\mu_{\rm B}$.
In particular, we find a strong in-plane anisotropy of $|A_a/A_{a^\ast} |\approx 4.1$.

Although the underlying hyperfine mechanism is not fully understood, the magnitudes of these constants are comparable to the direct dipolar couplings expected between V sites and localized Yb moments -- namely, $A_a^{\rm dip} \approx 21$ mT/$\mu_{\rm B}$, $A_{a^\ast}^{\rm dip} \approx -26$ mT/$\mu_{\rm B}$, and $A_c^{\rm dip} \approx 5$ mT/$\mu_{\rm B}$.
(These values are estimated by calculating the dipolar fields at the V sites from surrounding Yb$^{3+}$ moments, summed within a radius of $\sim$100 \AA.)
The observed in-plane anisotropy of $A_a/A_{a^\ast}$ appears to be greatly enhanced from the predicted rather isotropic dipole coupling ($|A_a^{\rm dip}/A_{a^\ast}^{\rm dip}| \approx 0.8$), suggesting an additional anisotropic contribution beyond the dipolar framework.
This may reflect an RKKY-type transferred hyperfine coupling mediated by conduction electrons, arising from the anisotropic character of the V-$d$ bands \cite{Pokharel2021Electronic-prop,Peng2021Realizing-Kagom}.

\subsection{Spin-lattice relaxation rate and spin dynamics}
\begin{figure}[bt]
\includegraphics[width=8.5cm]{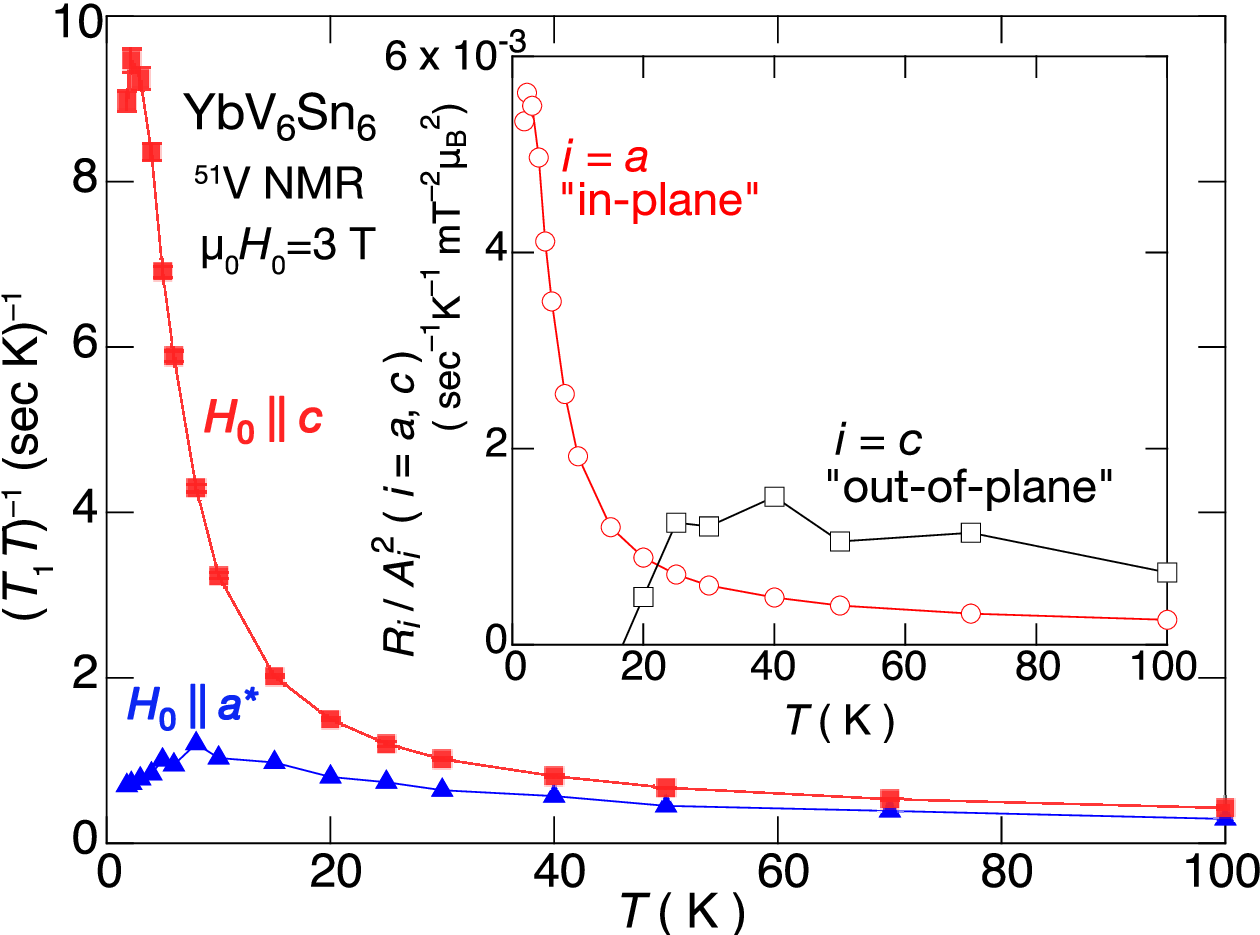}
\caption{\label{fig:invT1T_Ri} $^{51}$V NMR spin-lattice relaxation rate divided by temperature, $1/T_1T$, plotted against temperature, measured at $\mu_0 H_0 = 3$~T applied along the $c$ axis and $a^\ast$ axis.
Inset: Temperature dependence of $R_i/A_i^2$ representing the local spin fluctuations in the in-plane (along $i=a$ axis) and out-of-plane (along $i=c$ axis) directions (see the main text).}
\end{figure}

Spin-lattice relaxation rate ($1/T_1$) measurements were performed at $^{51}$V nuclei to further investigate the electronic properties of YbV$_6$Sn$_6$ from a local magnetic perspective. $T_1$ was recorded at $\mu_0H_0 =$~3~T
using the L3 line of the V1 site.
Relaxation curves were nicely fit with the form given in Sec.~\ref{sec:experimental} at all measured $T$,  confirming a spatially uniform relaxation that is governed by magnetic fluctuations. 
For most general case, $1/T_1$ can be written \cite{Moriya1963The-Effect-of-E} as
\begin{equation}
	\frac{1}{T_{1}} =  2 (\gamma_{\rm n}A_{\perp}/\gamma_{\rm e}\hbar)^{2}k_{\rm B}T\sum_{\bm{q}}f^{2}(\bm{q})\frac{{\rm Im}\chi_{\perp}(\bm{q}, \omega_{0})}{\omega_{0}},
\label{eq:invT1}
\end{equation}
where $\gamma_{\rm e}$ is the electronic gyromagnetic ratio, $A_i$ is the transferred hyperfine coupling constant, $f(\bm{q})$ is the hyperfine form factor (taken as unity for simplicity in the following analysis), ${\rm Im} \chi_{\perp}({\bm q}, \omega_{0})$ is the imaginary part of the transverse dynamic spin susceptibility of 4$f$ electrons, and $\omega_{0}$ is the nuclear resonant angular frequency.
Owing to the ${\rm Im} \chi_{\perp}$ term, $1/T_1$ picks up the low-energy spin fluctuations perpendicular to $H_0$.

Figure~\ref{fig:invT1T_Ri} shows the temperature dependence of $1/T_1$ divided by $T$, $(T_1T)^{-1}$, measured for $H_0 \parallel c$ and $H_0 \parallel a^*$.
$(T_1T)^{-1}$ is rather isotropic at high $T$ and shows a gradual increase with cooling in both field directions, consistent with the expected behavior for localized Yb 4$f$ moments.
Below approximately 20~K, however, a marked anisotropy sets in, and $(T_1T)^{-1}_{H_0 \parallel c}$ shows a pronounced increase toward lower $T$, whereas $(T_1T)^{-1}_{H_0 \parallel a^*}$ exhibits a broad maximum at a few Kelvin and then decreases upon further cooling, reaching $(T_1T)^{-1}_{H_0 \parallel c}/(T_1T)^{-1}_{H_0 \parallel a^*} \approx 13$ at 1.8~K.

To decompose in-plane and out-of-plane fluctuations, we define fluctuation rates along the $i$-axis as 
$R_i \equiv (\gamma_{\rm n}A_i/\gamma_{\rm e}\hbar)^2 k_{\rm B}\sum_{\bm{q}} {\rm Im}\chi_i(\bm{q}, \omega_{0})/\omega_{0}$ (with $i = a, a^\ast$, and $c$).
The fluctuation amplitude along the $i$-axis is then proportional to $R_i / |A_i|^2$.
Assuming that the in-plane anisotropy between the $a$ and $a^\ast$ directions is negligible, i.e., $R_a = R_{a^\ast}$, we obtain the following expressions:
$R_a = \frac{1}{2}(T_1T)_{H_0 \parallel c}^{-1}$ and $R_c = (T_1T)_{H_0 \parallel a}^{-1} - R_a$.
While this is a simplified approximation, we recognize that the in-plane anisotropy is actually important in this material, as indicated by the inequality $A_a \ne A_{a^\ast}$.
We aim to address this in future measurements of $1/T_1$ under $H_0 \parallel a$.

Despite its limitations, the approximation provides a useful first step for decomposing the spin fluctuations,
the inset of Fig.~\ref{fig:invT1T_Ri} shows the extracted fluctuation amplitudes along the $a$ and $c$ axes, $R_a / |A_a|^2$ and $R_c / |A_c|^2$, respectively, as functions of temperature.
From $(T_1T)^{-1}_{H_0 \parallel c}$ alone, we find that the in-plane fluctuations, represented by $R_a / |A_a|^2$, exhibit a pronounced enhancement below $\sim$20~K, becoming dominant at low temperatures.
Similar behavior has been reported in the Ce-based triangular-lattice antiferromagnet CeLi$_3$Bi$_2$ \cite{Bordelon2025Stripe-magnetic}.
In contrast, $R_c / |A_c|^2$ remains nearly constant above 20~K but decreases sharply at lower temperatures, becoming negative.
The apparent negative values below $\sim$20~K indicate a limitation of this simplified approximation and possibly reflect an anisotropic response of spin fluctuations to the applied field, with in-plane fluctuations selectively enhanced relative to the out-of-plane component.

Note that the temperature at which anisotropy of spin fluctuations significantly changes ($\sim$~20 K) approximately coincides with the reported energy separation between the CEF ground-state doublet and the first excited-state doublet, estimated as $\Delta/k_{\rm B} \approx 30$ K from magnetic susceptibility measurements~\cite{Guo2023Triangular-Kond}.
This suggests a possible change in the nature of the Yb 4$f$ moments due to thermal depopulation of the low-lying excited CEF level.
A subtle kink observed in the $K$–$\chi$ plot for $H_0 \parallel a$ (inset of Fig.~\ref{fig:KT_Kchi}) might be related to this change, as the largest hyperfine coupling in this direction could make such effects more detectable.
A similar anomaly is found in $^{23}$Na and $^{77}$Se NMR measurements in another Yb-based triangular-lattice system NaYbSe$_2$ \cite{Luther2024Anisotropic-mag}.

\section{Conclusion}

We report a detailed $^{51}$V NMR study of the vanadium kagome layers in the triangular Kondo lattice system YbV$_6$Sn$_6$ that exhibits magnetic order at $T_{\rm N} \approx 0.4$~K.
The single-crystal NMR spectrum under varying magnetic field directions determined the local EFG tensor and anisotropic hyperfine network of  $^{51}$V nuclei. 
Accurate alignments of single crystal enabled field-orientation dependent measurements of the spin-lattice relaxation rate $1/T_1$, which unveil a suppression of out-of-plane spin fluctuations below $\sim$20~K with a remarkable increase of in-plane fluctuations.
These results demonstrate that the low-$T$ electronic properties of this compound are primarily governed by in-plane spin dynamics.
The present findings highlight rich magnetic behavior in this layered $f$-$d$ electron system and provide a foundation for future investigations into field-induced quantum phases or magnetic anisotropy in related kagome-based Kondo lattice materials.
Further experiments are underway to explore the lower temperature properties of YbV$_6$Sn$_6$ in relation to the predominant in-plane spin fluctuations.

\begin{acknowledgments}
We thank P. F. S. Rosa for fruitful discussions.
Work at the Los Alamos National Laboratory, was primarily supported by the U.S. Department of Energy, Office of Basic Energy Sciences, Division of Materials Science and Engineering project ``Quantum Fluctuations in Narrow-band Systems".
H. S. and Y. T. acknowledge support from Japan Society for the Promotion of Science (JSPS) through KAKENHI (JP24KK0062 and JP23K25829).
S. H. acknowledges the Directed Funded Postdoctoral Fellowship through the Laboratory Directed Research and Development Program.
Work at the Japan Atomic Energy Agency (JAEA) was partially supported by the JAEA REIMEI Research Program.
Work at the Institute for Materials Research (IMR), Tohoku University was supported under the IMR-GIMRT program (Proposal Numbers 202406-HMKPB-0518, 202212-HMKBR-0507, 202312-HMKBR-0503).
\end{acknowledgments}


%

\end{document}